\documentclass[prl,aps,twocolumn,showpacs,preprintnumbers,amsmath,amssymb]{revtex4}
\usepackage[dvips]{graphicx}
\usepackage{amsmath}
\usepackage{amssymb}

\usepackage{graphicx}% Include figure files
\usepackage{dcolumn}% Align table columns on decimal point
\usepackage{bm}% bold math
\def\Dsl{\hbox{/\kern-.6700em\it D}} % D slash
\def\dsl{\hbox{/\kern-.5300em$\partial$}}

\def\eqa{\begin{eqnarray}}
\def\eeqa{\end{eqnarray}}
\def\eq{\begin{equation}}
\def\eeq{\end{equation}}
\def\be{\begin{equation}}
\def\ee{\end{equation}}
\def\bea{\begin{eqnarray}}
\def\eea{\end{eqnarray}}
\def\nn{\nonumber}

\begin{document}

\preprint{FERMILAB-PUB-06-365-A}

\title{A Note on Cosmic $(p,q,r)$ Strings}
\author{Mark G. Jackson}
\affiliation{Particle Astrophysics Center, Fermi National Accelerator Laboratory, Batavia IL 60510}

\date{\today}


\begin{abstract}
The spectrum of $(p,q)$ bound states of F- and D-strings has a
distinctive square-root tension formula that is hoped to be a
hallmark of fundamental cosmic strings. We point out that the Bogomol'nyi-Prasad-Sommerfield (BPS)
bound for vortices in ${\cal N}=2$ supersymmetric Abelian-Higgs
models also takes the square-root form. In contrast to string
theory, the most general supersymmetric field theoretic model
allows for $(p,q,r)$ strings, with three classes of strings rather
than two.  Unfortunately, we find that there do not exist BPS solutions
except in the trivial case.  The issue of whether there exist non-BPS solutions which may closely resemble the square-root form is left as an open question.
\end{abstract}

\maketitle

%%%%%%%%%%%%%%%%%%%%%%%%
\section{Introduction}
%%%%%%%%%%%%%%%%%%%%%%%%
There has been recent interest in the cosmic rehabilitation of
fundamental strings \cite{pol}. This has led to the exciting
possibility that the properties of a cosmic network made of
fundamental strings may have observationally distinct signatures
from more mundane solitonic objects. There are two smoking guns.
The first is the probability of reconnection which scales as
$P\sim g^2_s$ for fundamental strings \cite{jjp}, while it is
essentially unity for abelian vortices \cite{edneil,matz}.
Non-abelian vortices also typically reconnect with $P=1$ \cite{Eto:2006db}
although it is sometimes possible to get probability $P<1$ with the velocity dependence
substantially different from the fundamental string case \cite{mekoji}.

The second smoking gun \cite{cmp} \cite{Firouzjahi:2006vp} is the existence of both F- and
D-strings in warped IIB compactifications which form a distinctive
spectrum of bound states with tension
\begin{equation}
\label{ten}
 \mu_{(p,q)}=\sqrt{p^2\,\mu_F^2+q^2\,\mu_D^2}.
\end{equation}
A simple consequence of these bound states is the existence of
3-string junctions, with angles dictated by charge conservation
and the tension formula (\ref{ten}). Apects of network formation and
gravitational lensing of such junctions were studied in
\cite{edpaul,lens}.

It is rather simple to construct field theories which admit bound
states of vortices and the corresponding 3-string junctions.
Examples include vortices charged under discrete symmetries
\cite{discrete} and multiple abelian gauge groups \cite{paul}.
However, none of the field theoretic models studied so far
reproduce the stringy spectrum (\ref{ten}). The purpose of this
short note is to show that the general Bogomol'nyi bound in gauge
theories with multiple $U(1)$ gauge groups includes the string
spectrum (\ref{ten}). In fact, we shall see that there is a maximum
of three different types of supersymmetric vortices, with the
tensions bounded by
\be \mu =\sqrt{k_1^2 \mu_1^2+k_2^2 \mu_2^2+k_3^2 \mu_3^2}\label{tension}\ee
where $k_i$ are integer gauge winding charges.  If we choose a field theory without the third type of vortex, then this mimics the IIB string theory spectrum of cosmic strings.  Although we find that no nontrivial Bogomol'nyi-Prasad-Sommerfield (BPS) solutions exist which have this square-root spectrum, it is possible that non-BPS solutions exist which could closely resemble it.
%%%%%%%%%%%%%%%%%%%%%%%%
\section{The Bogomolnyi Bound}
%%%%%%%%%%%%%%%%%%%%%%%%
In theories with ${\cal N}=2$ supersymmetry, the real D-term and
complex F-term are unified into a triplet, transforming under an
$SU(2)_R$ R-symmetry. This existence of this triplet is
responsible for the three different tensions appearing in
(\ref{tension}). Recall that matter lives
in a hypermultiplet, consisting of two complex scalars $\phi$ and
$\tilde{\phi}$ transforming in conjugate representations of the
gauge group. For a single scalar charged under a single $U(1)$
gauge group, the D- and F-terms in the scalar
potential \footnote{There are further terms in the scalar potential
involving the vector multiplet scalar which may be consitently set
to zero.} are fixed by ${\cal N}=2$ supersymmetry to be,
\be
V=\frac{e^2}{2}(|\phi|^2-|\tilde{\phi}|^2-r_3)^2+\frac{e^2}{2}|2\tilde{\phi}
\phi-r_1-ir_2|^2.\ee
Here $e^2$ is the gauge coupling constant. There are three vacuum
expectation values $r_1$, $r_2$ and $r_3$ allowed by supersymmetry
(often referred to as a Fayet-Iliopoulos parameters). The
$SU(2)_R$ symmetry of this potential can be made manifest by
defining the doublet $\omega^T=(\phi,\tilde{\phi}^\dagger)$ and
writing
\be
V=\frac{e^2}{2}(\omega^\dagger\vec{\sigma}\,\omega-\vec{r})^2\ee
where $\vec{r}=(r_1,r_2,r_3)$ and $\vec{\sigma}$ are the triplet
of Pauli matrices.

Consider now a $U(1)^N$ gauge theory with gauge coupling $e^2_a, a= 1, \ldots,N$. We couple $N$ hypermultiplets $\omega_i$ with
integer charges $Q^i_{\ a}$ under the $a^{\rm th}$ gauge group.
The covariant derivatives are given by ${\cal
D}\omega_i=\partial\omega_i-i(\sum_{a=1}^NQ^i_{\ a}A_a)\omega_i$.
The energy functional for static $(\partial_0=A_0=0$)
configurations is
\be {\cal E}= \sum_{i=1}^N|{\cal D}\omega_i|^2+\sum_{a=1}^N\
\frac{1}{2e_a^2}B^2_a +\frac{e_a^2}{2}(\sum_{i=1}^N\,Q^i_{\
a}\omega_i^\dagger\,\vec{\sigma}\,\omega_i-\vec{r}_a)^2 . \label{energy}\ee
with $B_a$ the magnetic field for the $a^{\rm th}$ gauge group.
We choose $\det Q\neq 0$ to ensure that in the ground state,
defined by $\sum_i Q^i_{
\ a}\omega^\dagger_i\vec{\sigma}\omega_i=\vec{r}_a$, the $U(1)^N$
gauge group is fully broken and the theory exhibits a mass gap.

Lowest energy vortex states may be found by the usual Bogomolnyi
method. We search for straight strings, extended in the $x^3$
direction, by setting $\partial_3=A_3=0$ and writing
\begin{eqnarray*} && {\cal E} = \sum_{i=1}^N|{\cal
D}_1\omega_i-i(\vec{m}\cdot\vec{\sigma}){\cal
D}_2\omega_i|^2+ \\
&& \sum_{a=1}^N \frac{1}{2e_a^2} \left(\vec{m}B_a -
e_a^2
(\sum_{i=1}^N\,Q^i_{\ a}\omega_i^\dagger\,\vec{\sigma}\,\omega_i-\vec{r}_a)\right)^2
-B_a\vec{m}\cdot\vec{r}_a. \end{eqnarray*}
The above decomposition holds for any unit vector $\vec{m}$. The
last term yields a topological charge when integrated over the
plane transverse to the vortex string: $\int d^2x B_a=-2\pi k_a$.
Noting that the first two terms are squares, we derive
the bound on the tension
\begin{equation}
\label{bpstension}
\mu = \int d^2x\ {\cal E}\geq 2\pi \sum_a k_a\vec{m}\cdot
\vec{r}_a . 
\end{equation}
This is maximized by choosing $\vec{m}$ parallel to
$\sum_ak_a\vec{r}_a$.  In IIB string theory the tension-squared for a string with integer charge vector $k_a =(p,q)$ is expressed as 
\[ \mu^2 = \mathcal \sum_{a,b=1,2} (\mathcal M^{-1})^{ab} k_a k_b \]
where $\mathcal M_{ab}$ is the metric on the IIB auxiliary torus of modular parameter $\tau$.  We obtain the same spectrum by defining $(\mathcal M^{-1})^{ab} = 2 \pi \vec{r}_a \cdot \vec{r}_b$, where now $a,b =1, \ldots, N$.  In the special case where $(\mathcal M^{-1})^{ab} =  \mu^2_a \delta_{ab}$, this takes the form of (\ref{tension}).  Note that since the Fayet-Iliopoulos (FI) parameters $\vec{r}_a$ contain only three linearly independent directions, there are only three linearly independent effective string tensions, and we may henceforth assume that $N=3$ and so $k_a=(p,q,r)$.

The bound is saturated by solutions to the equations
\begin{eqnarray} \nonumber
\vec{m}B_a&=&e_a^2\left(\sum_{i=1}^N\,Q^i_{\ a}\omega_i^\dagger\,\vec{\sigma}\,\omega_i-\vec{r}_a\right) \\
\label{bog}
{\rm and} \ \ \ {\cal D}_1\omega_i &=&
i(\vec{m}\cdot\vec{\sigma})\,{\cal D}_2\omega_i
\end{eqnarray}
where, as explained above, $\vec{m}$ is the unit vector parallel
to $\sum_ak_a\vec{r}_a$. When all $\vec{r}_a$ lie parallel, for
example $\vec{r}_a=(0,0,r_a)$, these reduce to the usual coupled
vortex equations studied in \cite{mp}. They have solutions only
when the winding $n_i$ of all scalar fields with non-zero
expectation value, defined by $n_i=\sum_a Q^i_{\ a} k_a$ is
non-negative. (This is simply the statement that there is no
holomorphic vector bundle of negative degree). In this case there
is no attractive force between vortices. In contrast, when the
$\vec{r}_a$ do not lie parallel and the vortices in different
gauge groups are coupled through the scalars $\omega_i$, one may
expect bound states to form.

In both the field theoretic and string theoretic contexts, the
Bogomolnyi bound is expected to receive corrections at the scale
at which the protectorate supersymmetry is broken. In warped IIB
compactifications, supersymmetry is broken from 16 supercharges
(in the orientifold background) to 4 at the compactification scale,
with subsequent low-energy breaking at the TeV scale. 
%%%%%%%%%%%%%%%%%%%%%%%%
\section{Non-existence of BPS Solutions}
%%%%%%%%%%%%%%%%%%%%%%%%
Now decompose each field $\omega_i$ into
eigenvectors of $\vec{m}\cdot\vec{\sigma}$, writing
\be \omega_i=\psi_i \,|\vec{m}_+\rangle + \tilde{\psi}_i^\dagger
\, |\vec{m}_-\rangle . \ee
Then the covariant derivatives become
\begin{eqnarray*}
 {\cal D}_z\,\psi_i &\equiv&
(\partial_1-i\partial_2)\psi_i-i\sum_aQ^i_{\ a}(A^a_1-iA^a_2)\psi_i=0,
\nn\\ {\cal D}_{\bar{z}}\,\tilde{\psi}_i^\dagger &\equiv&
(\partial_1+i\partial_2)\tilde{\psi}^\dagger_i-i\sum_a
Q^i_{\ a}(A^a_1+iA^a_2)\tilde{\psi}_i^\dagger =0 .
\end{eqnarray*}
From this, we see that both $\psi_i$ and $\tilde{\psi}_i^\dagger$
transform with charge $Q^i_{\ a}$ under the $U(1)_a$ gauge group.
Taking the complex conjugate of the second equation, this
ensures that both $\psi_i$ and $\tilde{\psi}_i$ are covariantly
holomorphically constant, i.e.
\be {\cal D}_z\psi_i={\cal D}_z\tilde{\psi}_i=0 \label{bb1} \ee
where now $\psi_i$ has charge $Q^i_{\ a}$ while $\tilde{\psi}_i$ has
charge $-Q^i_{\ a}$. In other words, as the notation suggests, these
are the rotated form of $\phi_i$ and $\tilde{\phi}_i$. We can now
look at the first Bogomolnyi equation. Dotting with the unit
vector $\vec{m}$ tells us
\be B_a=e_a^2 \left(\sum_{i=1}^N Q^i_{\ a}
|\psi_i|^2-Q^i_{\ a}|\tilde{\psi}_i|^2-\vec{m}\cdot\vec{r}_a\right)\label{bb2} . \ee
Equations (\ref{bb1}) and (\ref{bb2}) are now in the form of the usual
coupled vortex equations described, for example, in Morrison and
Plesser \cite{mp}. The criterion for the existence of solutions is that for each scalar field $\psi_i$ we can define the
winding $n_i=\sum_a Q^i_{\ a}k_a$, while for each $\tilde{\psi}_i$ we
have $\tilde{n}_i=-\sum_a Q_{\ a}^ik_a$. Clearly $n_i=-\tilde{n}_i$.
Solutions to (\ref{bb1}) and (\ref{bb2}) exist if $n_i$ is non-negative
for each $\psi_i$ that gains an expectation value. (If $\psi_i$
has no expectation value for some $i$ then it may remain zero
throughout the solution). Similarly, each $\tilde{n}_i$ must be
non-negative for each $\tilde{\psi}_i$ which is non-zero. Clearly,
since $n_i=-\tilde{n}_i$, either $\psi_i$ or $\tilde{\psi}_i$ is
allowed an expectation value, but not both.

There are two further real equations that come from dotting the
first equation in (\ref{bog}) with $\vec{l}_\alpha$ where
$\vec{l}_{\alpha}\cdot\vec{m}=0$, for $\alpha=1,2$. We write
$\vec{l}=\vec{l}_1+i\vec{l}_2$. There is an ambiguous phase to the
vector ${\vec l}$, associated to rotating the basis $\vec{l}_1$
and $\vec{l}_2$, but we can always pick
a basis so that the remaining two real equations combine into the
complex equation
\be \sum_{i=1}^N Q^i_{\ a}\tilde{\psi}_i\psi_i=\vec{l}\cdot \vec{r}_a .\ee
Thus we see that in order for either $\psi_i$ or ${\tilde \psi}_i$ to be zero, the vector $\sum_a (Q^{-1})^i_{\ a} {\vec r}_a$ must be perpendicular to ${\vec l}$, making it proportional to ${\vec m}$.  Since ${\vec m} \propto \sum_a k_a {\vec r}_a$, this requires $k_a$ to be proportional to $(Q^{-1})^i_{\ a}$ (and of course $k_a$ must be integer-valued).  That is, in order for only $n_i$ to be nonzero we select $k_a$ to be the $i$th entry of $Q^{-1}$.  There is also the trivial solution when all $\vec{r}_a$ lie parallel, so that each $\vec{r}_a$ is proportional to ${\vec m}$ and thus perpendicular to $\vec{l}$, but this does not produce a square-root spectrum.

While these are the necessary conditions for BPS solutions, we find they are not sufficient.  Consider the radially symmetric field ansatz (a non-radially symmetric solution would necessarily have higher energy and thus could not be BPS):
\be \omega_i=\left(\begin{array}{c}\psi_i \\
\tilde{\psi}^\dagger_i
\end{array}\right)=\left( 1 - q_i (\rho) {\vec m} \cdot {\vec \sigma} \right) \left(\begin{array}{c} s_i \\ \tilde{s}_i \end{array}\right)\, e^{in_i \theta}\ee
where we use polar coordinates $(\rho,\theta)$ in the $(x^1,x^2)$-plane.  The asymptotic boundary condition is
\be q_i \rightarrow 0\ \ \ {\rm as}\ \ \
\rho\rightarrow \infty \nn \ee
making the vacuum selection at infinity
\begin{equation}
\label{vac}
\sum_iQ^i_a\
(s^*_i,\tilde{s}^*_i)\,\vec{\sigma}\,\left(\begin{array}{c} s_i
\\ \tilde{s}_i\end{array}\right) = \vec{r}_a . 
\end{equation}
The ansatz for the gauge potential is
\be A^a_\mu =
\frac{\epsilon_{\mu \nu} x^\nu}{\rho^2} A_a(\rho), 
\hspace{0.5in} A_a = - k_a + \rho f_a(\rho) \ee
where we similarly require $f_a \rightarrow 0$
as $\rho\rightarrow \infty$.  The field strength is given by the simple expression
\be B^a = \partial_x A^a_y - \partial_y A^a_x = \frac{1}{\rho} \frac{\partial A_a}{\partial \rho}. \ee
To determine the long-distance behavior of the fields, we insert the ansatz into the equations (\ref{bog}) and expand to linear order in $q_i$ and $f_a$ to obtain
\begin{eqnarray}
\label{lin1}
 &  \left( \frac{f_a}{\rho} + f'_a \right) = -2 e_a^2 \sum_iQ^i_a (|s_i|^2+| {\tilde s}_i|^2) q_i,& \\
\nonumber
& \left(  {\vec m} \cdot {\vec \sigma} \partial_\rho - \sum_a Q^i_a f_a \right) \left( 1 - q_i (\rho) {\vec m} \cdot {\vec \sigma} \right) \left(\begin{array}{c} s_i
\\ \tilde{s}_i\end{array}\right) = 0 . &
\end{eqnarray}
The second equation can be solved (to linear order!) to give
\begin{equation}
\label{lin2}
q'_i = - \sum_a Q^i_{\ a} f_a .
\end{equation}
Differentiation of these first-order equations then produces the modified Bessel equations
\begin{eqnarray*}
f''_a + \frac{1}{\rho} f'_a - \frac{1}{\rho^2} f_a - \sum_b L^2_{ab} f_b &=& 0, \\
q''_i + \frac{1}{\rho} q'_i - \sum_j M^2_{ij} q_j &=& 0 
\end{eqnarray*}
where the mass-squared matrices are given by
\begin{eqnarray*}
 L^2_{ab} &=& 2 e^2_a \sum_i (|s_i|^2 + |{\tilde s}_i|^2)  Q^i_{\ a} Q^i_{\ b}, \\
 M^2_{ij} &=&  2 (|s_i|^2+| {\tilde s}_i|^2) \sum_a e_a^2  Q^i_{\ a} Q^j_{\ a}.
 \end{eqnarray*}
As should be expected from a BPS solution, the gauge and matter mass-squared matrices $L^2_{ab}$ and $M^2_{ij}$ have identical eigenvalues, which can be seen by acting with $Q^i_{\ a}$ as a similarity transformation.  Denoting the mass eigenvalues as $\lambda_A$ (so that the mass-squared eigenvalues are $\lambda^2_A$), the solution to (\ref{lin1}) and (\ref{lin2}) is then given by
\begin{eqnarray}
\nonumber
f_a &=& \sum_A S_{a A} C_A \lambda_A K_1 (\lambda _A \rho), \\
\label{asym}
q_i &=&  \sum_{a,A} Q^i_{\ a} S_{a A} C_A K_0 (\lambda_A \rho)
\end{eqnarray}
where $S_{a A}$ is the diagonalization matrix for $L^2_{ab}$ and the coefficients $C_A$ cannot be determined in the linear approximation and would have to be fixed from numerical comparison to the non-linear solution.  We would expect that only gauge fields charged under the given $k_a$ and the matter field $\omega_i$ with nonzero winding $n_i$ should attain a profile, but from (\ref{asym}) we see that the $Q^i_{\ a}$ mix the matter and gauge fields into a basis such that non-charged fields are excited.  To prevent this it must be that the $L^2_{ab}$ and $M^2_{ij}$ are diagonal, making the $C_A$ proportional to $k_a$.  This gives us a total of 6 constraints (3 each from setting the off-diagonal components of a symmetric matrix to zero).  These are precisely enough constraints to set the off-diagonal components of $Q^i_{\ a}$ to zero, which now makes the interaction trivial.  The acceptable BPS windings are then simply of the form $k_a = (p,0,0), (0,q,0)$ or $(0,0,r)$, which will not display any distinctive square-root behavior.  The lack of BPS solutions for such $N=2$ theories has been noted previously in the literature \cite{Lee:2005sv} \cite{Eto:2005sw} but not in the context of cosmic $(p,q,r)$ strings.

The situation is very reminiscint of supersymmetric quantum mechanics, with the doublet $\omega_i=(\psi_i, {\tilde \psi}_i^\dagger)$ taking the place of the bosonic and fermionic components of the wavefunction.  The ground state (BPS) solution (when it exists) is given by first-order solutions which allow one component or the other, but not both.  It is still likely that non-BPS solutions exist which will contain both components, but it will have energy greater than the BPS bound.  

\section{Conclusion}
We have shown the BPS spectrum for supersymmetric vortices exhibits the same square-root cosmic string spectrum as superstring theory, including not just two but three types of vortices.  Unfortunately no BPS solutions exist which actually exhibit this square-root spectrum.  It is still likely that non-BPS solutions exist which would have an energy higher than the BPS bound, but which might approximate the square-root BPS spectrum for a certain choice of parameters.  It would be interesting to make a full analysis of these solutions.

\section*{Acknowledgements} I would like to thank D. Tong for suggesting this idea and much assistance during its completion, and M. Strassler for useful discussions.  I am supported by NASA grant NAG 5-10842.

\end{document}